\documentclass{webofc}

\usepackage[varg]{txfonts}   
\usepackage{hyperref}
\usepackage{url}
\hypersetup{colorlinks=true,citecolor=blue,urlcolor=blue,linkcolor=blue}
\begin{document}
\title{Jet Modification and Medium Response - Theory Overview}
%
%

\author{\firstname{Daniel} \lastname{Pablos}\inst{1}\fnsep\thanks{\email{daniel.pablos@usc.es}}
}

\institute{Instituto Galego de F\'isica de Altas Enerx\'ias IGFAE, Universidade de Santiago de Compostela,
E-15782 Galicia-Spain}

\abstract{This text contains a summary and personal perspective on the current status and challenges of jet quenching physics as portrayed by the presentations delivered at the 12th International Conference on Hard and Electromagnetic Probes of High-Energy Nuclear Collisions (Hard Probes 2024) which took place in September 2024 in Nagasaki, Japan. 
}
\maketitle

\section{Introduction}
\label{intro}
Jets are collimated sprays of hadrons that are the result of the evolution in virtuality of energetic partons produced in hard scatterings. As they propagate through a deconfined QCD medium, such as the quark-gluon plasma (QGP) formed in heavy-ion collisions at the LHC and RHIC, many of their properties are heavily modified with respect to those measured in proton-proton collisions. These set of phenomena, commonly referred to as \emph{jet quenching}, allow us to infer unique information about the QGP by means of model comparison to experimental data. The most prominent and well established of these modifications consist in the depletion of high-$p_T$ jet yields~\cite{ATLAS:2023iad}, accompanied by an excess of soft particles around the jet axis~\cite{ATLAS:2019pid}. These observations are consistent with what one expects from energy loss calculations due to the presence of deconfined QCD matter both at weak~\cite{Blaizot:2012fh} and strong coupling~\cite{Chesler:2015nqz}. By energy-momentum conservation, it is expected that as the jet is modified by the medium, the medium gets modified by its interaction with the jet as well. This backreaction of the medium is known as \emph{medium response}, and is mainly modelled using two different frameworks. For high-enough momentum transfers via elastic scatterings with the medium constituents, one needs to consider the dynamics of the recoiling particles, which can further reinteract with the medium. These physics probe the short-length, perturbative structure of the QGP. For those lower momentum transfers of non-perturbative nature, or directly because of the thermalization of softer jet partons, the excitation of hydrodynamic wakes is expected, thereby probing the long wavelength features of the QGP.

A crucial step forward that has been undertaken in the last years has consisted in the extension of energy loss calculations from the static QGP brick setup to more realistic scenarios where the dynamical nature of the medium is accounted for. Indeed, finding experimental evidence of the spatiotemporal evolution of the flowing QGP using jet observables is one of the key goals of jet quenching physics. In this conference we have witnessed exciting developments in this direction, such as new calculations on the effects of local medium flow and gradients on stimulated gluon radiation~\cite{Kuzmin:2023hko,XMayo,Barata:2024bqp,JSilva,CSalgado} as well as predictions of drift effects causing jet deflection in preferred orientations~\cite{Bahder:2024jpa,JBahder,RFries,TLuo}. Naturally, one also expects the evolution of jet-induced wakes to be affected by medium dynamics~\cite{Tachibana:2015qxa}. We need to hunt for these effects in real data to fully exploit the tomographic potential of jets, 
and in this way obtain further evidence to corroborate and complete our current understanding of the fluidlike QGP evolution.

\section{Jet Evolution}
It is well known that a jet is not simply a set of on-shell charges. From vacuum physics one has learned that jets are better understood as a collection of color correlated dipoles. In the medium, energetic jets also develop a high-virtuality, vacuum-like evolution that interacts with that medium. To date, the interplay between vacuum and medium scales still remains a great theoretical challenge, one that is necessarily affected by color coherence physics.

The paradigmatic example in vacuum involves the successive soft gluon emissions off a partonic dipole. The wavelength $\lambda$ of the soft emission can only resolve the presence of the two color charges of the dipole if it is smaller than their distance $d$ at the formation time of the emission $\tau_f$. This condition immediately implies that the angle of the soft emission has to be smaller than the opening angle of the dipole, $\theta<\theta_{q\bar{q}}$, for the soft gluon to be effectively emitted from one or the other leg of the dipole. Following this reasoning down to further emissions in the parton shower results into the so called angular ordering phenomenon, which is the basis for the probabilistic Monte Carlo approach to jet evolution.

\begin{figure*}
\centering
\vspace*{1cm}       
\includegraphics[width=6.4cm,clip]{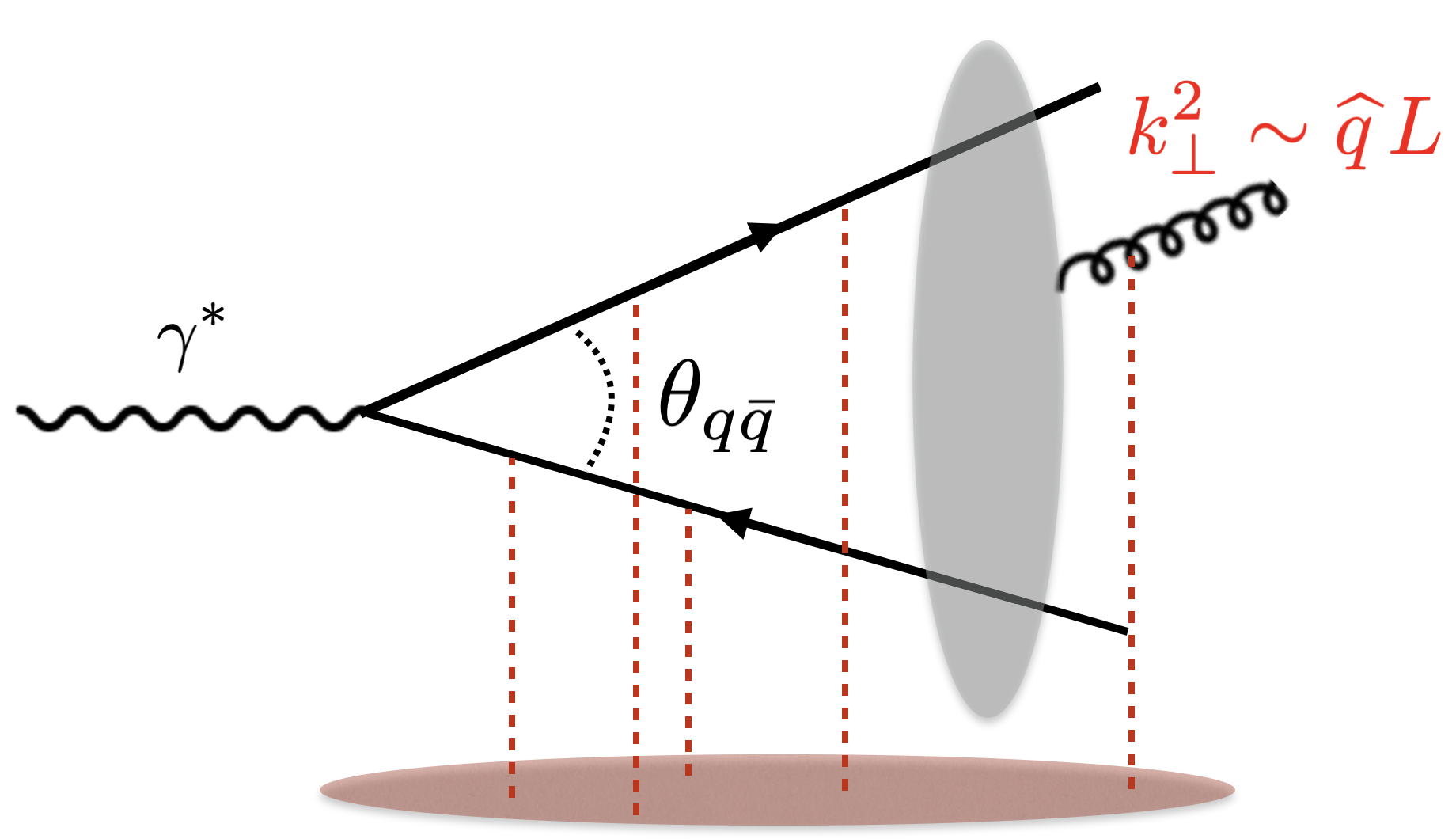}
\includegraphics[width=6cm,clip]{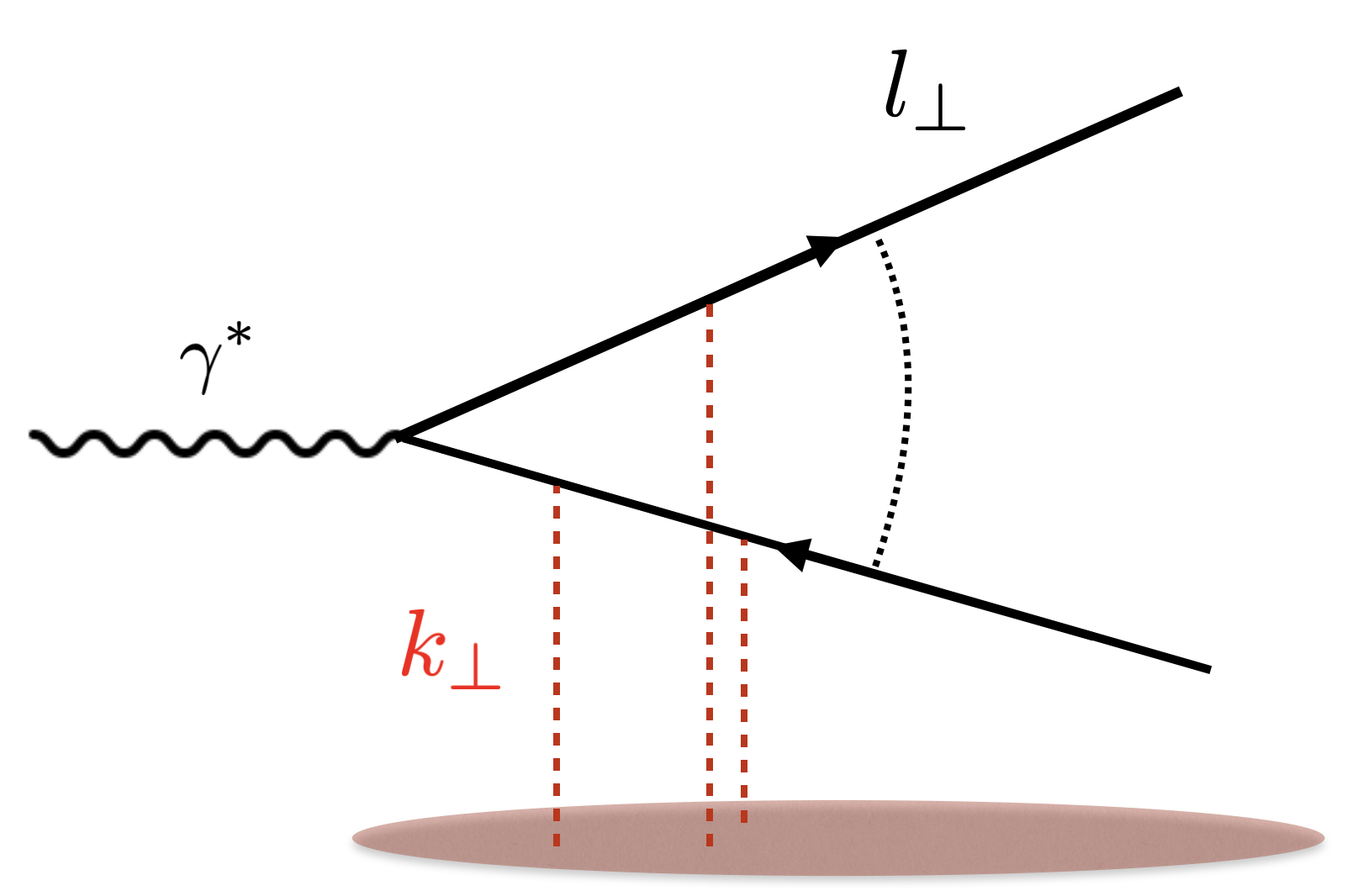}
\caption{Sketches of color singlet dipoles, produced by high-virtuality jet evolution, which interact with the medium (red blob). \emph{Left}: Stimulated gluon emission via multiple soft scatterings during time $L$ possess an off-shellness of order $k_{\perp}^2\sim \hat{q} L$. For the emission to be able to resolve the individual dipole charges, the angle of the dipole needs to be $\theta_{q\bar{q}}>\theta_c\sim 1/\sqrt{\hat{q}L^3}$. \emph{Right}: A given exchange $k_{\perp}$ with the medium can only resolve a short-lived high virtuality dipole with transverse momentum $l_{\perp}$ if its wavelength is small enough, so $k_{\perp}>l_{\perp}$. This fact restricts potential modifications of the early stages of the shower to come from the rarer large momentum exchanges.}
\label{fig-1}       
\end{figure*}

In the medium, a similar simplified reasoning allows us to understand the role played by color coherence in jet quenching. The wavelength of a stimulated emission via multiple soft scatterings, with transverse momentum $k_{\perp}^2\sim \hat{q}L$, in a medium of length $L$, can only resolve the individual charges at the maximum possible formation time $\tau_f\sim L$ if the angle of the antenna is larger than some critical coherence angle, i.e. $\theta_{q\bar{q}}>\theta_c\sim1\sqrt{\hat{q}L^3}$ (see sketch in the left panel of Fig.~\ref{fig-1}). This means that a dipole with an angle smaller than $\theta_c$ will only produce medium-induced emissions off its total charge. Clearly, obtaining an increasingly precise understanding of color coherence physics in the medium is indispensable for a more accurate phenomenology. So far, the leading effects have been implemented in a number of models~\cite{Hulcher:2017cpt,Caucal:2019uvr,Mehtar-Tani:2024jtd}, showing substantial impact in virtually every observable.

Coherence effects also play a role in the potential modifications of the high-virtuality stage of jet evolution. Not any momentum transfer can resolve a short-lived highly virtual dipole with transverse momentum $l_{\perp}$; only those with transfers $k_{\perp}>l_{\perp}$ will (see sketch in the right panel of Fig.~\ref{fig-1}). Since high-momentum transfers are relatively rare, this implies that the effective $\hat{q}$ decreases with increasing virtuality of the dipole~\cite{Kumar:2019uvu}. The inclusion of this so-called modified coherence effects results in the mitigation of the impact of medium-modified splitting functions during the early vacuum-like evolution of the jet. It has been shown that these considerations improve the simultaneous description of single inclusive hadron and single inclusive jet suppression~\cite{JETSCAPE:2022jer,AKumar,PJacobs} in the MATTER+LBT setup within JETSCAPE. When not included, the excess of stimulated radiation off the leading particle in the jet translates into a value of $R_{AA}$ that is too low for single inclusive hadrons. The importance of the interplay between the vacuum and medium scales has also been highlighted by a recent study using MARTINI~\cite{Modarresi-Yazdi:2024vfh,SShi}, where it is shown that accounting for the finite formation time of vacuum-like splittings greatly improves the description of several jet observables.

A very convenient way to study the modifications in the jet shower due to medium interactions is by using the Lund plane, which maps the density of splittings inside the jet as a function of their angle $\theta$ and transverse momentum $k_{\perp}$. In particular, performing cuts at different energy scales $k_{\perp}$ allows us, to a certain extent, to disentangle the role played by the different physical mechanisms~\cite{Cunqueiro:2023vxl}. The study of the angular distribution of those splittings above a certain high $k_{\perp}$-cut precisely probes the potential modifications during the high-virtuality stage, while lower values of $k_{\perp}$ are more sensitive to coherence physics and the presence of semi-hard recoils from medium response. The measurement, which is underway, has the potential to experimentally access the scale down to which a vacuum-like jet evolution develops within the medium. The usage of the Lund plane is natural as well within more formal contexts aimed at developing a framework for jet quenching based on Effective Field Theory techniques, as outlined in recent efforts~\cite{Mehtar-Tani:2024smp,YMehtar}, where one exploits the scale separation between the vacuum-like hard-collinear modes undergoing DGLAP evolution and the medium-modified collinear-soft modes that are encapsulated in Wilson line correlators.

\section{Medium Response}
We have seen a number of examples concerning the physics of color coherence and jets within the medium, and so it is natural to ask what are the role played by these effects in the phenomenon of medium response. Typically, one considers that elastic scatterings between jet partons and the medium constituents happen as if the jet parton was an on-shell particle coming from infinity. However, we know that they are in general part of a dipole. So, as long as the dipole has not been decohered due to multiple soft scatterings, a semi-hard elastic scattering cannot be said to have been triggered by one or the other leg of the dipole, and quantum interference effects have to be taken into account. This simple consideration has potentially wide phenomenological applications, and the first steps have been taken in~\cite{Pablos:2024muu}. It is shown that in the limit in which the energy of the recoiling medium parton (a quark) is much softer than the energy of the dipole, but much larger than the medium particle rest mass, its angle is constrained to be smaller than the opening angle of the dipole (for a color singlet dipole), very much the same as in the soft gluon emission in vacuum example discussed above. 
Additionally, this means that the collisional energy loss of a dipole actually depends on its substructure, as $\hat{e}\equiv dE/dx \sim \log \theta_{q\bar{q}}$. This effect is the first known relation between color coherence and medium response, and is yet to be implemented in models.

The dynamics of recoils are believed to be of great importance for jet quenching phenomenology. We have recently seen that they could represent the largest contribution to the large angle enhancement measured by CMS in
the energy-energy correlator (EEC) observable~\cite{CMS:2025ydi,JViinikainen}, according to models that include both medium-induced radiation and recoils, such as LBT~\cite{Yang:2023dwc,ZYang,Xing:2024yrb,WXing} and JEWEL, opening up new ways with which to constrain the scales at which medium response physics dominate. Among several remarkable properties, the peak position of the EEC does to some extent remember the scale at which the jet was produced, enabling a procedure to remove some selection bias effects~\cite{Andres:2024hdd,CAndres} and thus allowing a more direct determination of the genuine medium-induced modifications we are after.

\begin{figure*}
\centering
\vspace*{1cm}       
\includegraphics[width=6.4cm,clip]{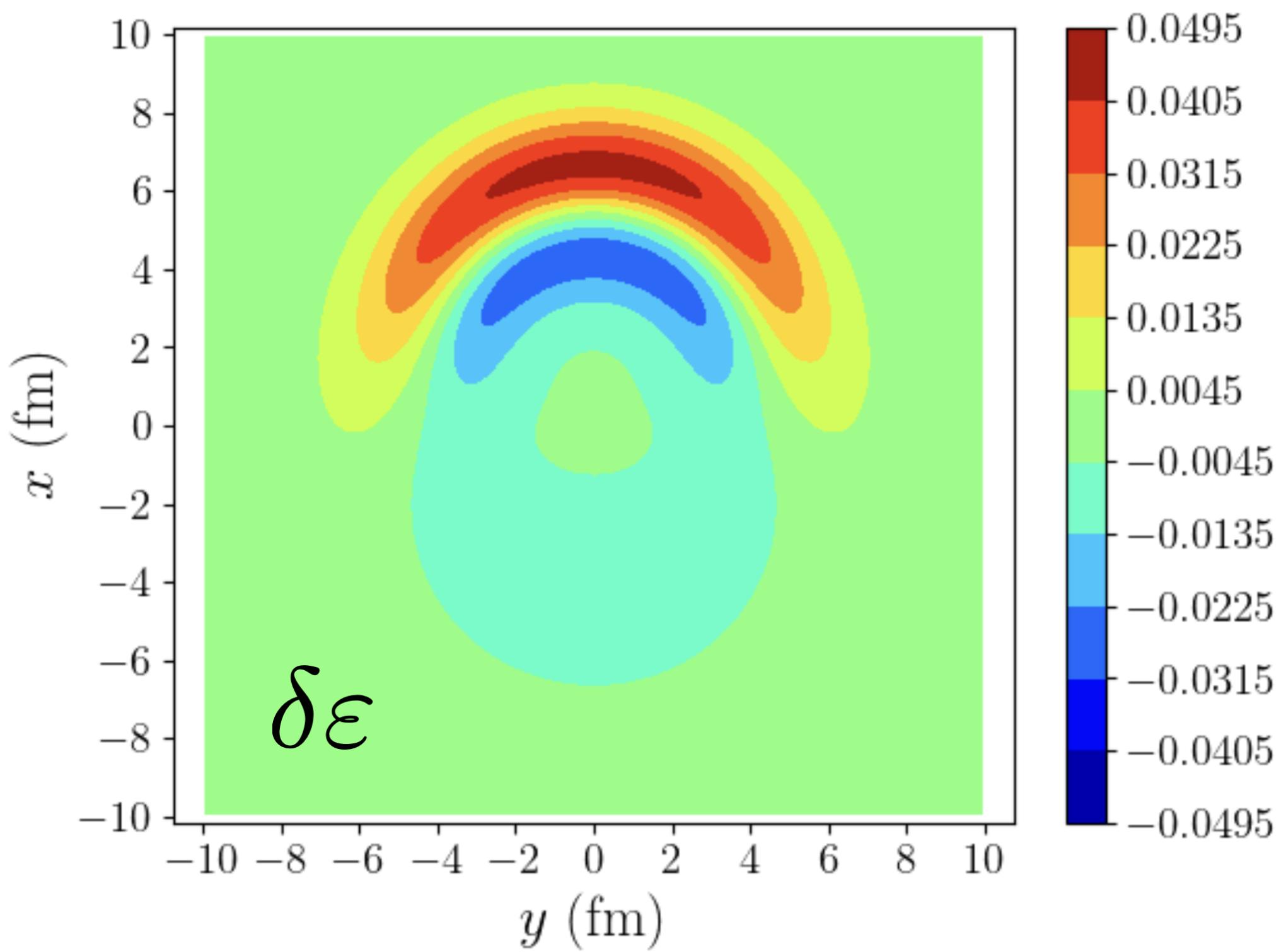}
\includegraphics[width=6.4cm,clip]{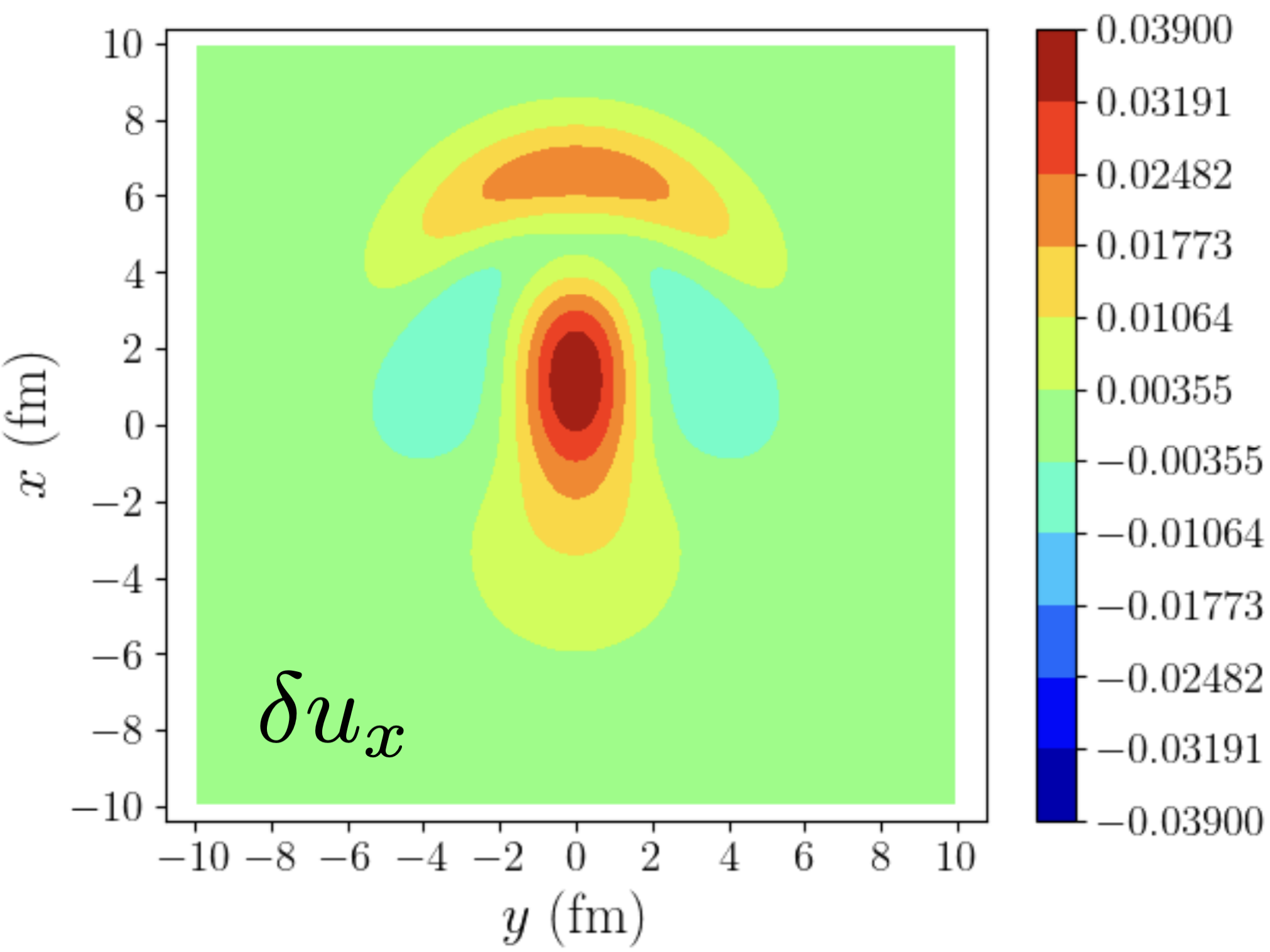}
\caption{Transverse profile of the fluid perturbations induced by the deposition of energy and momentum in a viscous QGP by an energetic quark, after 6 fm/c of the last deposition. \emph{Left}: energy density perturbations $\delta \varepsilon$, mostly corresponding to sound modes, with a Mach cone-like wavefront. They carry little momentum and lead to isotropic particle production at freezeout time. \emph{Right}: velocity perturbations $\delta u_x$ in the direction of propagation of the energetic quark. They mostly correspond to the diffusion wake contribution, which carries most of the momentum. They lead to an excess of particles in the jet direction with a corresponding depletion in the opposite direction at freezeout time (with respect to an unperturbed background). Figures taken from~\cite{Casalderrey-Solana:2020rsj}.
}
\label{fig-2}       
\end{figure*}

The other way in which medium response effects manifest themselves is by the excitation of hydrodynamic wakes due to the jet passage through the flowing QGP. They arise due to the injection of energy and momentum into the medium by means of a source term in the hydrodynamic equations of motion. Perturbations in energy density $\delta \varepsilon$ are propagated in the form of sound waves, which for a supersonic jet result in a Mach cone structure (see left panel of Fig.~\ref{fig-2}). These modes carry very little momentum. Momentum perturbations (proportional to) $\delta u$ along the direction of the jet result in the so-called \emph{diffusion wake} structure, which is the most phenomenologically relevant since it carries most of the momentum. These modes describe moving fluid that stays behind the jet, where the perturbation was produced, undergoing a recirculating motion possessing vorticity (see right panel of Fig.~\ref{fig-2}).

The impact of the hydrodynamic wake on most observables requires the usual mapping between fluid variables and particles degrees of freedom using the Cooper-Frye procedure at the freezeout hypersurface. It is easy to see that the contribution coming from the sound waves produces soft particles with a fairly isotropic angular distribution, while the contribution from the diffusion wake results in an excess of soft hadrons in the direction of the jet accompanied with a depletion in the opposite direction (in the transverse plane) with respect to an unperturbed fluid (the uncorrelated background typically subtracted in experiments). In this context, the excess and depletion (with respect to the uncorrelated background) can be understood from the boost experienced by those fluid cells affected by the momentum injected by the jet -- while in the fluid cell rest frame particle production is isotropic, boosting back to the laboratory frame introduces preferred orientations according to the direction and magnitude of the velocity of the fluid cell.

Naturally, this excess and depletion relative to the jet orientation is also observed in Effective Kinetic Theory implementations that account for energy-momentum conservation, such as~\cite{He:2015pra,Mehtar-Tani:2022zwf}. The elastic scatterings between a jet parton and QGP constituents cause the replacement of a thermal parton moving in a random direction (in a static medium) with a recoiling parton that is preferentially moving along the jet direction. Overall, this leaves a net excess of soft partons along the jet direction together with a depletion in the direction opposite to it.

In this conference we have seen new proposals with which to study the dynamics of jet induced hydrodynamic wakes. The 3-point energy correlator (EEEC) observable, consisting in the energy-weighted angular distribution of triplets of particles within a jet, presents a striking dependence on the presence of the wake, which is specially manifest in the equilateral region of the triangle~\cite{Bossi:2024qho}. This is a fairly unpopulated region in vacuum, since it is away from the small angle region dominated by the QCD collinear singularity, shaped as squeezed triangles. While the EEC observable can identify the angular scale of the underlying dynamics, the EEEC (and higher point correlators) could be used to map out the dynamics themselves. 

Another new proposal revolved around the idea of analyzing multiple wakes. Over the last years, we have learned that jet structures which undergo a more active fragmentation during its vacuum-like evolution tend to be more suppressed than those with a less active, and so narrower, fragmentation pattern. The question is then whether one can analyze the features of the wakes that arise due to multiple structures being quenched. While it would be tremendously complicated to do this at the level of individual partons within a typical jet, the new procedure pioneered by ATLAS~\cite{ATLAS:2023hso,MRybar} to reconstruct large-R jets using solely skinny small-R jets makes this idea possible. By studying the angular distribution of soft particles within the large-R jet as a function of the angular separation of the two skinny subjets, one is actually looking at the shape of two wakes with different degree of overlap~\cite{Kudinoor:2025ilx,AKudinoor}. This opens up the possibility to experimentally study their interference patterns as well as their coupling with the background flow depending on their orientation with respect to the event plane of the collision.

\section{The Smoking Gun}

While it seems clear that many observables that present an excess of soft particles around the jet direction are affected by the physics of medium response, it is also true that medium-induced radiation similarly leads to a turbulent cascade of soft quanta that extends rapidly to large angles as well. The extent to which these two mechanisms of soft particle production lead to similar qualitative pictures is well exemplified by the so-called missing-$p_T$ observable by CMS~\cite{CMS:2015hkr}. This very differential observable dissects the momentum imbalance between the leading and subleading jet hemispheres as a function of the angle with respect to the dijet axis, and is binned in the $p_T$ of the charged tracks. Both the turbulent cascade picture~\cite{Blaizot:2015lma} and the hydrodynamic wake profile~\cite{Casalderrey-Solana:2016jvj} present a qualitatively satisfactory agreement with experimental data, in particular regarding the need to go to fairly large angles to restore momentum balance in the dijet system.

What is truly unique about medium response, i.e. absent in other mechanisms, is the depletion in the direction opposite to the drag experienced by the fluid. The first experimental explorations started around 5 years ago using boson-jet systems. The main advantage of boson-jet systems is that the colorless boson (a photon or a $Z$) recoiling from the quenched jet does not lose energy to the plasma, and therefore the boson hemisphere does not contain any potential wakes, or other quenched or unquenched hadronic structures that could overlap with the signal: a yield depletion in the boson hemisphere (opposite to the quenched jet in the transverse plane) with respect to the uncorrelated background. The first searches using $Z$-hadron correlations~\cite{CMS:2021otx} did not correct for the contamination coming from multi-parton interactions. More recent searches using photon-jet ensembles~\cite{ATLAS:2024prm,YGo} suffer from limited sensitivity, likely due to the imposition of a jet $p_T$ cut, although the existence of a measured depletion is not ruled out with current uncertainties.

A new (preliminary) measurement by CMS~\cite{CMS:2024fli,YJLee} presented at this Hard Probes conference in Nagasaki holds promise to become the first unambiguous evidence of the QGP response to the jet passage. The azimuthal angle distribution of charged hadrons with respect to the $Z$-boson shows a clear enhancement in the (unmeasured) jet direction, as well as a \emph{clear depletion in the $Z$ direction}. The experimental results are confronted with four different models with fairly different physical ingredients. Only those models that include medium response (either the hydrodynamic wake or recoils from elastic scatterings) succeed in reproducing the data. Furthermore, the depletion found in the $Z$ hemisphere is seen to be located around the rapidity of the $Z$, which is again well reproduced only by models that include medium response. The pull suffered by the QGP in the jet direction means that the wake should actually be oriented around the azimuthal angle and rapidity of the (unmeasured) jet. However, due to kinematical constraints (i.e. triggering on a fairly boosted $Z$) both the $Z$ and the recoiling jet tend to lie around midrapidity. 
Unique diffusion wake effects can also be revealed using the much more abundant dijet samples, where the overlap between the two wakes can be engineered by setting different rapidity gaps between the dijet system~\cite{Pablos:2019ngg}.

In a really complex field such as heavy-ion collisions it is not so often that one encounters clearly distinctive qualitative effects that are intimately related to a specific physical mechanism. While from a theoretical point of view it can be argued that medium response has to be there, simply due to energy-momentum conservation, it is unclear a priori whether these effects can actually be measured in experiments, or whether they are overwhelmed by other effects. If confirmed, this new $Z$-hadron correlations measurement represents experimental evidence of the phenomenological importance of medium response. Even more, by adjusting models to get the right size of the measured depletion in the $Z$ side, one also gets the size of the associated enhancement due to medium response in the jet side, thereby constraining the available room for potential contributions from alternative particle production mechanisms.

\section{Conclusions}

The advent of new calculations that quantify the imprints of the flowing QGP properties on jet radiation and broadening brings us closer than ever to the true era of jet tomography. Ongoing developments in our understanding of the interplay between the vacuum and medium scales are crucial to provide the complete picture. These include the determination of the potential modifications of early vacuum-like splittings as well as improved precision in the description of color coherence effects. If confirmed, the new CMS measurement on $Z$-hadron correlations can represent a milestone in jet quenching physics. Medium response physics would then be of experimentally proven importance for phenomenology. It is evident that further work is needed to consistently put together the dynamics of recoils with that of hydrodynamic wakes in models. A truly meaningful interpretation of observables will come from models including state-of-the-art implementations of both medium-induced radiation an medium response so as to make robust statements about the nature of QGP via the physics of jet quenching.

\section*{Acknowledgements}
DP is funded by the European Union's Horizon 2020 program under the Marie-Curie grant agreement No 101155036 (AntScat), by the European Research Council project ERC-2018-ADG-835105 YoctoLHC, by Spanish Research State Agency under project PID2020-119632GB- I00, by Xunta de Galicia (CIGUS Network) and the European Union, and by Unidad de Excelencia Mar\'ia de Maetzu under project CEX2023-001318-M.
%
%

\end{document}